# In connection with identification of VLF emissions before L'Aquila earthquake

M. K. Kachakhidze[1], Z. A. Kereselidze[2], N. K. Kachakhidze[1]

[1] St. Andrew The First-Called Georgian University of The Patriarchy of Georgia, Tbilisi, Georgia

[2] Iv. Javakhishvili Tbilisi state university, M. Nodia Institute of Geophysics, Tbilisi, Georgia

*Correspondence to:* M. K. Kachakhidze    manana_k@hotmail.com

**Abstract.** The present paper deals with an attempt to check up the theoretical model of self-generated seismo-electromagnetic oscillations of LAI system on the basis of retrospective data.

Application of the offered simple model enables one to explain qualitatively the mechanism of VLF electromagnetic emission initiated in the process of an earthquake preparation.

It is worth to pay attention to the fact that frequency changes from MHz to kHz in electromagnetic emission spectrum comes to a good agreement with avalanche-like unstable model of fault formation.

L'Aquila earthquake taken as an example to isolate reliably the Earth VLF emission from the magnetospheric electromagnetic emission of the same frequency range, MHD criterion is offered together with geomagnetic activity indexes.

On the basis of the considered three earthquakes, according to the opinion of authors the model of self-generated seismo-electromagnetic oscillations of the LAI system will enable us to approach the problem of resolution of earthquake prediction by certain accuracy.

## 1 Introduction

The present study is based mainly on the data of analysis of Abruzzo earthquake, given in Special Issue of "Natural Hazards and Earth System Sciences" (NHESS - Special Issue, 2010). In this Issue there are described all phenomena accompanying L'Aquila catastrophic earthquake occurred on 6 April 2009.

In our paper we aim to check up the feasibilities of application of our schematic model (Kachakhidze et al., 2011) with the view of reliability of earthquakes prediction.

According to the offered model a segment of the Earth where tectonic stress in earthquake preparation period anomalously increases should have positive electric potential towards atmosphere. The presence of such effect is proved in the papers (Bleier, et al, 2009; Eftaxias et al, 2009).

It is known that the period of strong earthquake preparation is rather long, although some effects associated with it are manifested only when the tectonic stress value approaches a limit of geological environment solidity.

Abundant laboratory, ground and satellite observations proved that "fracture-induced physical fields allow a real-time monitoring of damage evolution in materials during mechanical loading. When a material is strained, electromagnetic (EM) emissions in a wide frequency spectrum ranging from kHz to MHz are produced by opening cracks, which can be considered as so called precursors of general fracture" (Bahat et al., 2005; Eftaxias et al., 2007a; K. Eftaxias et al, 2009, Hayakawa and Fujinawa, 1994; Hayakawa, 1999; Hayakawa and Molchanov, 2002).

It should be noted that the stressed rock behaves like a stress-EM transformer. These phenomena are detectable both at laboratory and geological scale (Hayakawa and Fujinawa, 1994; Hayakawa, 1999; Hayakawa et al., 1999; Gershenzon and Bambakidis, 2001; Hayakawa and Molchanov, 2002; Bahat et al., 2005; Eftaxias et al., 2007a; Muto et al., 2007; Hadjicontis et al., 2007). Studies on the small (laboratory) scale reveal that the MHz EM radiation appears earlier than the kHz one, while the kHz EM emission is launched from 97% up to 100% of the corresponding failure strength (Eftaxias et al., 2002). On the large (geological) scale, intense MHz and kHz EM emissions precede EQs that: (i) occurred in land (or near coast-line), (ii) were strong (magnitude 6 or larger), (iii) were shallow (Eftaxias et al., 2002, 2004, 2006, 2007b; Kapiris et al., 2004a, b; Karamanos et al., 2006). Their lead time is ranged from a few days to a few hours. Importantly, the MHz radiation precedes the kHz one at geophysical scale, as well (Eftaxias et al., 2002; Kapiris et al., 2004a; Contoyiannis et al., 2005). Notice that a complete sequence of SES, MHz and kHz EM anomalies have been observed one after the other in a series of significant EQs that occurred in Greece (Eftaxias et al., 2000, 2002; Karamanos et al., 2006; Papadimitriou et al., 2008). K. Eftaxias notes as well: "We argue that the detected MHz and kHz EM radiation anomalies were emitted from the focal area of the L'Aquila EQ (Eftaxias et al., 2010). Scientists pay attention to the fact that "The time lags between preearthquake EM anomalies and EQs are different for different types of precursors. The remarkable asynchronous appearance of precursors indicates that they refer to different stages of EQ preparation process" (Eftaxias et al., 2010).

It is considerable that theoretical model (Kachakhidze et al, 2011) doesn't contradict above described observed facts.

## 2 Discussion

### 2.1 How strong will the incoming earthquake be

The formula (1) (Kachakhidze, et all., 2011) has been checked up on the basis of retrospective data with the view of experimental evaluation of a theoretical model. The formula connects with each other analytically the main frequency of the observed electromagnetic emission and the linear dimension (the length of the fault) of the emitted body:

$$\omega = \beta \frac{c}{l} \qquad (1)$$

where $\beta$ is the characteristic coefficient of geological medium and it approximately equals to 1. Of course it should be determined individually for each seismically active region, or for a local segment of lithosphere.

According to formula (1), for instance, when the frequency is $10^5$ Hz, fault length $l$ already equals to 3 km. It is clear that further decrease of frequency will refer to increase of the fault length, and hence to the expected strong earthquake.

Thus, the formula (1) leads us to the purposeful monitoring of electromagnetic emission spectrum that will enable us to follow the process of formation of the main fault.

Our model enables us to describe successively the precursory effects taking the catastrophic earthquake of L'Aquila as an example.

According to observed data of L'Aquila ULF, MHz and kHz EM anomalies were detected prior to the L'Aquila EQ that occurred on 6 April 2009. The initially emerged MHz EM emission is thought to be due to the fracture of a highly heterogeneous system that surrounds a family of large high-strength asperities distributed along the fault sustaining the system. The finally emerged strong impulsive kHz EM radiation is due to the fracture of the asperities themselves (Eftaxias, et al, 2010).

It is known that electromagnetic anomalies in MHz were fixed on March 26 2009 and April 2 2009 (Eftaxias et al, 2010). According to our model it implies that in the vicinity of the incoming earthquake focus the formation of cracks was started, the length of which varied within the frames 300 meter.

Electromagnetic anomalies were observed in kHz on April 4 2009 (Eftaxias et al, 2010) – referring to the fact that the fault length in the focus was already of the kilometer order.

According to our model in the process of uniting of cracks and formation of the main fault, the polarization charges which are formed of the fault surface finally are arranged along the main fault. At this moment separate cracks might play the role of thermoionized channels with different electric conductivity.

Total length of a thermoionized channel of high electric conductivity, with the geological point of view, is the main fault length in the focus and its relation to VLF of EM emission is expressed by the formula (1).

According to the offered model, and correspondingly, the formula (1), it is evident that when fracture stops, the emitted EM emission sieges, which is confirmed by experimental data (Eftaxias et al., 2010).

To demonstrate the adequacy of the theoretical model and the experimental data, alongside with the earthquake of Italy, we have considered other cases of earthquakes (China, Haiti):

1) Wenchuan earthquake (May 12, 2008, $M$ 8.0, depth $\sim 19$ km., 31.121N,103.367E).

It is known that the main frequency of the electromagnetic emission prior to the considered earthquake equaled to approximately 1 kHz, which, according to the formula (1), when $\beta = 1$, corresponds to the fault length $l \approx 300$ $km$ in the earthquake focus. By the use of this parameter we can calculate a theoretical magnitude. With this in view, we used Ulomov's formula (Ulomov., 1993), which is just for $M \geq 5.0$ magnitude earthquakes:

$$\lg l = 0,6 \ M - 2,5 \qquad (2),$$

that is

$$M \approx 8,1$$

which practically coincides with real magnitude of this earthquake (Xuemin Zhang, 2010).

2) L'Aquila earthquake (April 6, 2009, $M$ 6.3, depth $\sim$10 km, 42.35N 13.38E).

According to the data available for this earthquake the fault length in the focus $l \approx 15$ $km$ (Cirella et.al, 2009). Before the earthquake the main frequency of electromagnetic emission was approximately 20 kHz (Rozhnoi, et all., 2009). For this frequency the formula (1) offers the same value of a fault length, while the magnitude calculated according to the Ulomov's formula coincides with the value of real magnitude of the earthquake.

3) Haiti earthquake (January 12, 2010, $M$ 7.0, depth $\sim$ 13 km, 18,443 N, 72.571W).

In case of Haiti earthquake it was impossible to determine the fault length in the focus. It appears that there was a superimposition of a new fault over the old one.

Our model enables us to resolve roughly the inverse problem too. Namely: due to the fact that M=7.0 for this earthquake (Athanasiou et all, 2011) according to the Ulomov's formula we'll have:

$$l \approx 50 km$$

which, in case of $\beta = 1$, according to the formula (1) will conform to:

$$\omega \approx 6 \text{ kHz}$$

It is known that in this case electromagnetic emission, in frequency range (0–20 kHz), were recorded by the satellite DEMETER, concerning a time period of 100 days before and 50 days after this earthquake.

Thus, we can take about 10 kHz as the main characteristic frequency, for which $l \approx 30 \; km$.

Thus, the offered model enables us to adjust, gradually, step by step, the magnitude of the forecasted earthquake by means of monitoring of electromagnetic emission of the period preceding the earthquake.

## 2.2 When will an earthquake occur?

The model offered by us is just for shallow inland earthquakes of the magnitude $M \geq 5.0$. Pursuant to the Ulomov's formula, for such magnitudes, fault length in earthquake focus is already of a kilometer order.

According to the formula (1) appearance of kHz in frequency spectrum means that the main fault starts its formation, but future frequency decrease refers to the increase of incoming earthquake magnitude.

It should be stated that in the first approximation, the analysis of the experimental data proves our theoretical conclusions (Kachakhidze et al., 2011). It is also known from the experiments that strong earthquake occurs in one or two days after kHz appears in electromagnetic emission (Eftaxuas, et al., 2010, Papadopoulos, et al., 2010).

Generally, $\omega(\beta)$ frequency of emission depends on the properties of a geological medium. Therefore, if we know the characteristic $\omega(\beta)$ frequency for concrete seismic region, determined on the basis of the retrospective data, the electromagnetic emission monitoring in the period preceding the earthquake will enable us to forecast the time of earthquake occurring with certain accuracy.

## 2.3 Where will an earthquake occur?

According to our model, it is possible to detect the territory on the surface of the Earth in advance, where an earthquake is expected - the epicentral area of a future earthquake will be approximately limited to the territory where the earth surface will have positive potential towards atmosphere. Electromagnetic emission in kHz should take place namely on the territory adjoining the epicenter of a future strong earthquake. Our opinions have been experimentally confirmed by the paper of K.Eftaxias (Eftaxias et al., 2010).

Alongside with it, papers were published in Special Issue (NHESS – Special Issue, 2010) in which various researchers consider the problem of a location of an incoming earthquake on the basis of experimental data (Rozhnoi et al., 2009; Biagi et al., 2009). These works are in good conformity with our theoretical model.

## 2.4 Avalanche-like unstable model of fault formation with the view of the theoretical model of self-generated seismo-electromagnetic oscillations of LAI system

It should be emphasized that the avalanche-like unstable model of fault formation that is well known in seismology (Mjachkin et al., 1975) coincides with the model offered by us (Kachakhidze et al) and explains well the given succession of EM emission changes in time taking place in the period of earthquake preparation: MHz, kHz and disappearance of emission directly before an earthquake (Johnston 1997, Eflaxias et al, 2009).

It is known that avalanche-like unstable model of fault formation is divided into three stages (Fig. 1): the first stage, which for strong earthquakes can last for several ten months and is not considered the so-called "precursory stage", because during this stage the chaotic formation of only micro cracks without any orientation occurs.

This stage of formation of microcracks is reversible process - at this stage not only microcracks can be formed but also their the so-called "locked" can occur. Cracks created at this stage will be small (some dozen or hundred meter order) and therefore the first stage in the electromagnetic emission frequency range, according to our model (Kachakhidze et al., 2011) should be expressed by the discontinuous spectrum of MHz order emission, which is proved by the latest special scientific literature (Eftaxias K., et al., 2009; Papadopoulos G., et al., 2010).

The second stage of the avalanche-like unstable model of fault formation is an irreversible avalanche process of already somewhat oriented microstructures, which is accompanied by inclusion of the earlier "locked" sections. We have to suppose that this stage in the emission frequency spectrum should be expressed by MHz continuous spectrum already. According to the avalanche-like unstable model, this process takes place some 10-14 days before an earthquake. Observations over earthquakes prove vividly that in electromagnetic emission frequency spectrum the continuous emission appears in MHz range namely 10-14 days before an earthquake (Papadopoulos G., et al., 2010).

According to the avalanche-like unstable model, at the very stage gradual increase of cracks occurs (up to the kilometers order) at the expense of their uniting, to which, according to our model, from the formula (1) corresponds to the transition of MHz to kHz emission in the electromagnetic emission frequency spectrum.

At the final, third stage of the avalanche-like unstable model of fault formation the relatively big size faults use to unite into one, the main fault. This process, according to our model, in case of emission spectrum monitoring should correspond to gradual fall of frequencies in kHz, which according to the formula (1) refers to the increase of fault length in the focus.

Final formation of the main fault, that is relatively unstable zone, is accompanied by total fall of average "macro" stress in the greater part of the volume, since at this stage new faults are not created any more. This effect, according to our model, should be expressed in disappearance of EM emission that really takes place some hours before an earthquake. This fact is proved by observations on earthquakes (Johnston, 1997, Eftaxias, et al, 2009). The next, fourth stage in the avalanche-like unstable model of fault formation corresponds to the moment of an earthquake occurrence.

It should be emphasized that monitoring of EM emission frequency spectrum enables us to distinguish rather simply series of foreshocks and aftershocks from the main shock.

**2.5 Magnetospheric VLF electromagnetic emission**

Practically such emission always accompanies the increase of solar wind gasdynamics pressure. The same effect occurs in the case of connection of the interplanetary magnetic field on the magnetosphere day-side and the geomagnetic field border force line as well.

This effect, as a rule, incites perturbation of inner structures of the magnetosphere and often it is crowned with the global geomagnetic storm. At this moment the Earth VLF emission will surely be overlapped by a far more intensive magnetospheric emission (Akhoondzadeh et al, 2009). Therefore, in the seismically active middle latitudinal belt the Earth VLF electromagnetic emission should be reliably isolated from the magnetospheric VLF emission of the same frequency range.

Namely, in the process of analysis of the Earth VLF emission nobody supposed a possibility of generation of magnetospheric VLF electromagnetic emission in calm, or less perturbed geomagnetic conditions. It is quite natural that application of Earth VLF emission for the purposes of earthquake prediction would not be correct without clarification of a problem.

There is a model (Kereselidze et al, 1993), which aims to demonstrate the possibility of generation of VLF electromagnetic emission in the middle latitudes, at the calm and less perturbed geomagnetic conditions. Fundamental basis of this diagnostic model is the classical analytical solution obtained by S. Chapligin, which is just for ideal, incompressible liquid stream approximation. Application of this solution will enable one to simulate the process of flow of the magnetosphere dayside by solar wind (Aburdjania et al, 2007). Large-scale hydrodynamic picture corresponding to it contains the so-called stagnation zone. Maximum dimension of this formation is conditioned by the thickness of the magnetosheath preceding the magnetosphere and the speed of solar wind in interplanetary space.

It is considered that at certain conditions, even at less perturbed solar wind the large-scale laminar stream of solar wind plasma in the magnetosheath preceding dayside of magnetosphere passes to turbulent one. The magneto hydrodynamic criterion of the maximal size stagnation zone border stability can be used as an indicator of changes of magnetosphere flow regime

$$M = 1 - v_1/2v_a \geq 0 \qquad (3)$$

where $v_1$, $v_a$ - are the solar wind speed and Alphen speed in plasma at the border of stagnation zone correspondingly.

Methodology of calculation of (3) criterion for maximal size stagnation zone is rather simple (Kereselidze et al, 1993), and the necessary data for any moment, are accessible, on the corresponding cites (http://swdcdb.kugi.koyotee.ac.jp).

According to formula (3), the mark "plus" corresponds to a large-scale laminar regime of magnetosphere flow, while "minus" – to a turbulent one. In the last case the number of solar wind energetic electrons, which enter magnetosphere from the bottom of the stagnation zone, might suffer increase, and their stream might contribute to initiation of electron cyclotron instability in plasmosphere (ionosphere) captured plasma and to the generation of the VLF electromagnetic emission.

This idea was checked rather reliably some time ago (Kereselidze et al., 1993). It turned out that in calm and less perturbed geomagnetic conditions there the morphological relation really exists between the magnetosphere flow regime and magnetospheric VLF electromagnetic emission. The middle latitudinal VLF emission data were used at the level of upper ionosphere together with the indexes indicating the level of geomagnetic activity. Global middle latitudinal geomagnetic index satisfies the condition $D_{st} \geq -15\,\text{nT}$ (nanotesla), while the global magnetic activity index $K_p \leq 2$ corresponds with the sufficient reserve, to the calm or less perturbed magnetosphere.

The analysis showed that when M $\geq 0$, that is, when according to diagnostic model the solar wind flow large-scale structure near the magnetosphere border had to be laminar, the VLF emission was not fixed practically. Instead, VLF emission was almost always fixed when $M < 0$, that is, probably at the turbulent regime of solar wind flow. Aggregate prediction accuracy of these results was rather high ($\approx 85\%$). Alongside with it, further studies revealed that at the increase of latitudinal interval the diagnostic possibilities of the given model decrease. This fact once more refers that plasmosphere should be the source of the magnetospheric VLF electromagnetic emission (Nishida, 1978). Thus, we consider rather reasonable that estimation of geomagnetic activity by $K_p$ and $D_{st}$ indexes (Akhoondzadeh et al. 2010) is not sufficient for the exclusion of the possibilities of generation of magnetospheric VLF emission. With this in view, application of additional M criterion (index) makes rather strict the demand of geomagnetic indexes, which is proved by the below given example.

Figures 2-4 offer dynamic picture of changes of the $K_p$, $D_{st}$ indexes and M criterion in three month interval covering L'Aquila earthquake. Geomagnetic perturbation level is limited to the values $D_{st} \geq -15$ and $K_p \leq 2$ (on figures the moment of earthquake occurring is marked by a black dot). We have to state that the Fig. 2 practically coincides with the data of the above referred paper (Akhoondzadeh et al. 2010) (where for $D_{st}$ index a relatively less strict term is used: $D_{st} \geq -20$ nT), while the picture corresponding to $K_p$ is identical with our picture.

Fig.2. Changes of $K_p$ index in time interval corresponding to L'Aquila earthquake;

Fig.3. Changes of $D_{st}$ index in time interval corresponding to L'Aquila earthquake;

Fig.4. Changes of M index in time interval corresponding to L'Aquila earthquake.

Simple analysis shows that often, when according to $K_p$, $D_{st}$ indexes magnetosphere might be considered calm or less perturbed, M criterion is negative. It implied that the magnetospheric VLF electromagnetic emission may exist at middle or lower latitudes.

## Conclusions

The model of self-generated seismo-electromagnetic oscillations of the LAI system somewhat simplifies consideration of various problems associated with earthquakes:

1. For reliable separation of the Earth VLF electromagnetic emission in the seismic active middle latitudinal belt from the same frequency range magnetospheric electromagnetic emission, the special MHD criterion has been offered.

2. EM emission earlier considered as earthquake indicator only, turned out an earthquake precursor because it "brings" very precious information for prognostic diagnostic of the incoming earthquake.

3. In case of electromagnetic emission spectrum monitoring in the period that precedes earthquake it is possible to determine, with certain accuracy, the time, location and magnitude of an incoming earthquake simultaneously.

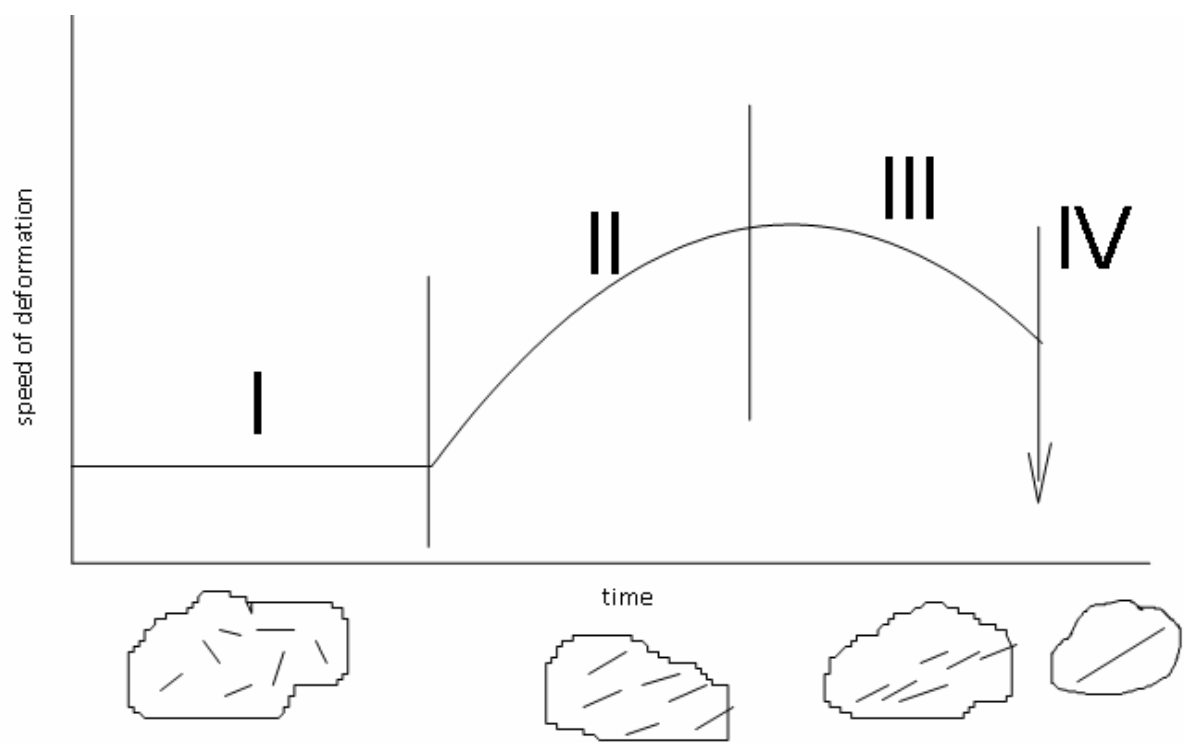

Fig.1. Scheme of fall-unstable model of fracture origination

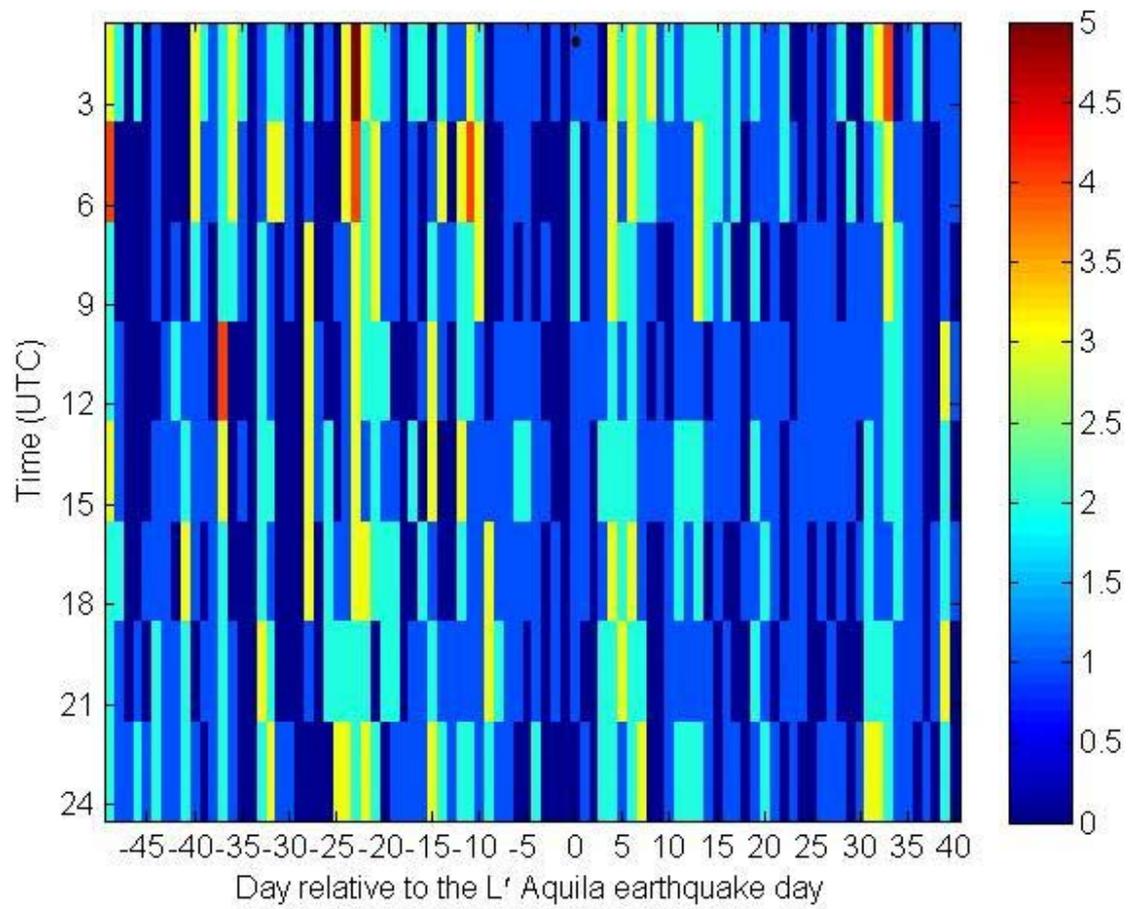

Fig.2. Changes of $K_p$ index in time interval corresponding to L'Aquila earthquake

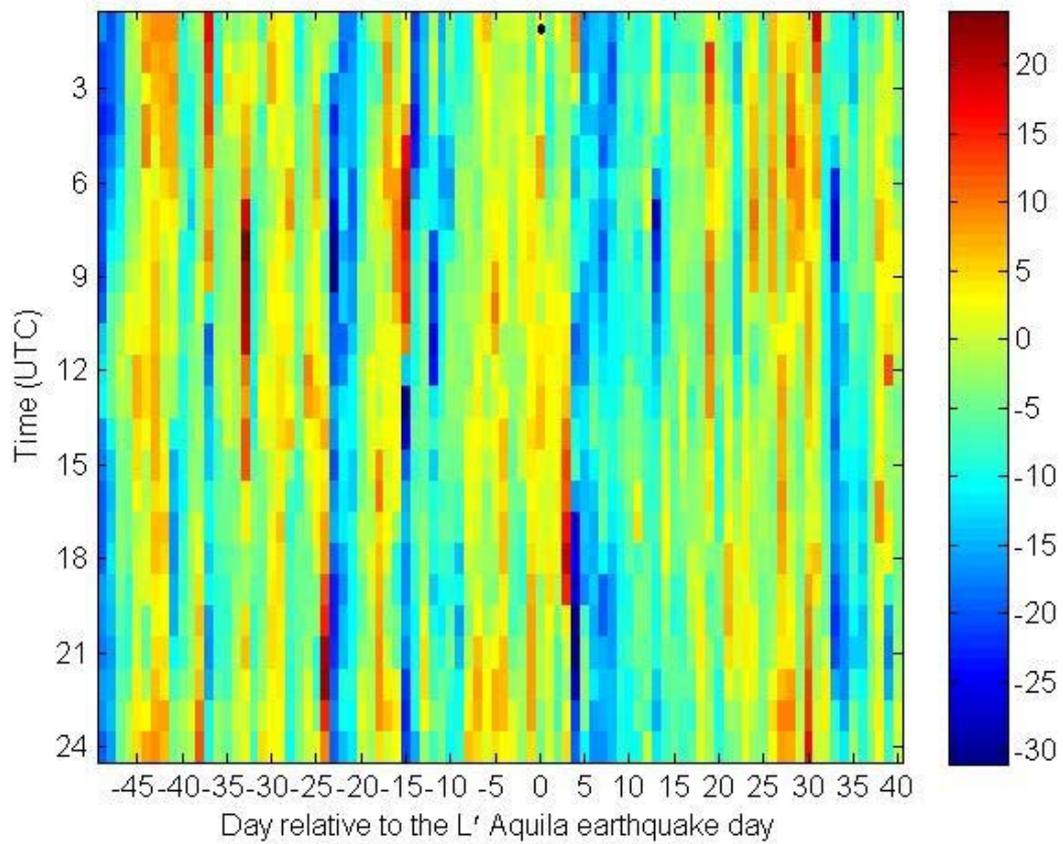

Fig.3. Changes of $D_{st}$ index in time interval corresponding to L'Aquila earthquake

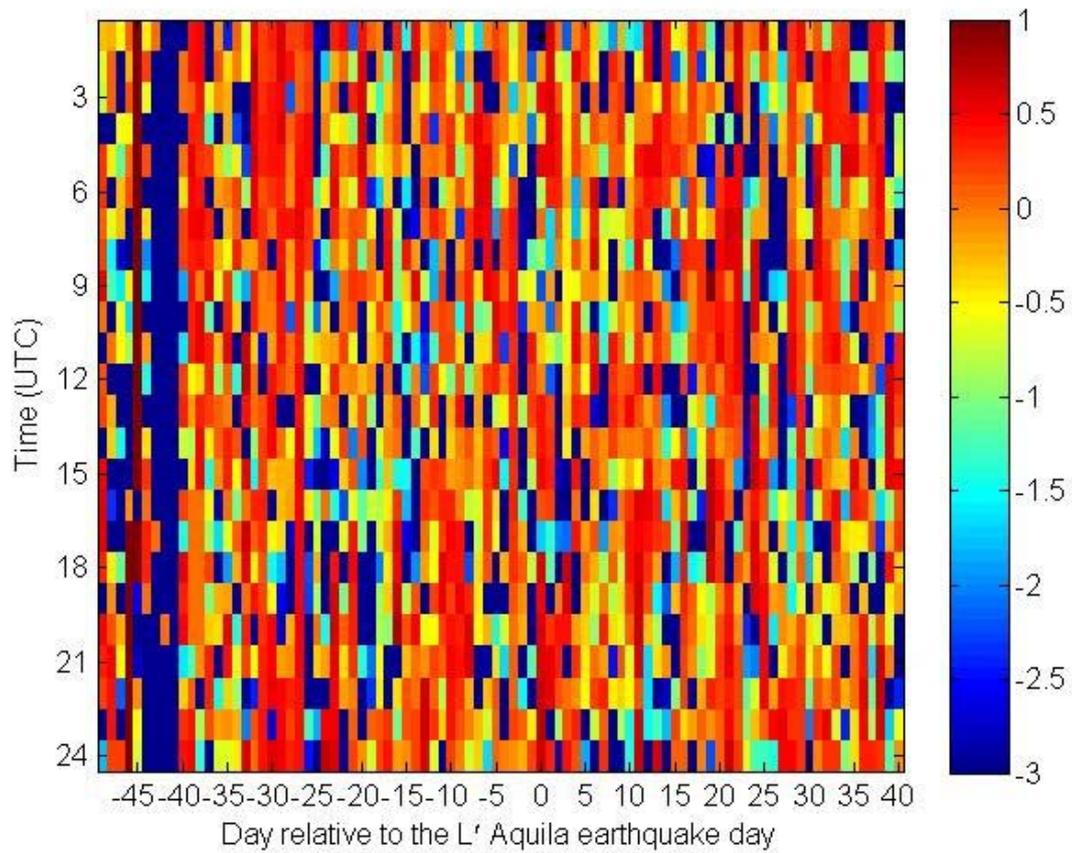

Fig.4. Changes of M index in time interval corresponding to L'Aquila earthquake